\documentclass[prb,twocolumn,superscriptaddress,showpacs,amsmath,amssymb]{revtex4-1}

\usepackage{graphicx,color}
\usepackage{subcaption}
\usepackage{amsmath}
\usepackage{amssymb,bm,bbm}

\definecolor{MS-color}{RGB}{128,0,128}

\begin{document}

\title{
 Magnetic field-controlled 0-$\pi$ transitions and their experimental signatures in superconductor-ferromagnet-superconductor junctions
}

 \author{A. Vargunin}
 \affiliation{Department of
Physics and Nanoscience Center, University of Jyv\"askyl\"a, P.O.
Box 35 (YFL), FI-40014 University of Jyv\"askyl\"a, Finland}
\affiliation{Institute of Physics, University of Tartu, Tartu, EE-50411, Estonia}
 \author{M. A. Silaev}
 \email{mikesilaev@gmail.com}
 \affiliation{Department of
Physics and Nanoscience Center, University of Jyv\"askyl\"a, P.O.
Box 35 (YFL), FI-40014 University of Jyv\"askyl\"a, Finland}

\affiliation{Moscow Institute of Physics and Technology, Dolgoprudny, 141700 Russia}
\date{\today}
\begin{abstract}
Superconductor-ferromagnet-superconductor Josephson junctions are known to 
exist in the $0$ and $\pi$
states with the transitions between them controlled by the temperature and ferromagnetic interlayer thickness. 
We demonstrate that these transitions can be controlled also by the external magnetic field directed perpendicular to the layers. By varying the ratio of diffusion coefficients in superconducting and ferromagnetic layers, these field-controlled transitions can be made detectable for arbitrary large value of the exchange energy in the ferromagnet. 
We 
also show that the $0$-$\pi$ transitions in the perpendicular field can be observed as the specific features 
of the  flux-flow conductivity dependencies on the ferromagnetic thickness in accordance with recent experimental results. 
 \end{abstract}
\maketitle

\begin{figure}[h!]
\includegraphics[width=0.9\linewidth,trim={1cm 0.5cm 3cm 0.5cm},clip]{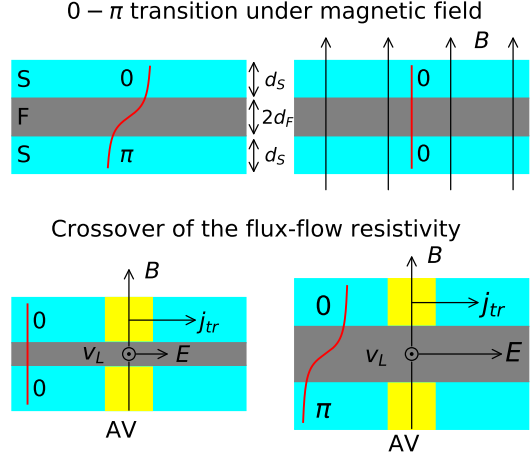}
 \caption{\label{f:sfs} 
 Upper panels: SFS structure where the increasing  perpendicular magnetic field $\bm B$ results in the transition from $\pi$ to  $0$-state. 
  Lower panel: Crossover of the flux-flow resistivity $\rho_{ff}$ of the SFS structure upon the transition from $0$ to $\pi$
  state with the change of 
  F layer thickness. 
  $\bm 
  v_L$ is the  velocity of Abrikosov vortex (AV) shown by the yellow color, $\bm j_{tr}$ is the transport current, $\bm E = \rho_{ff}\bm j_{tr}$ is the electric field.  }
\end{figure}

Superconductor-ferromagnet-superconductor (SFS)  junctions such as shown schematically in Fig. \ref{f:sfs} are known to have either $0$ or $\pi$ ground state Josephson phase difference\cite{buzdin1982,buzdin1991}. Switching between these states controlled by  
the parameters such as temperature $T$ and ferromagnetic interlayer thickness $d_F$ are governed by the 
oscillations of the 
Cooper pair wave functions as a result of the energy splitting between the spin-up and spin-down states introduced by the exchange field $h$ \cite{Buzdin2005,lyuksyutov2005}.

Transitions between the 0 and $\pi$-states 
have been observed experimentally as the strong oscillations of the critical current of a junction 
\cite{ryazanov2001,kontos2002,shelukhin2006observation,oboznov2006}. The $\pi$ state can be also revealed 
in the closed electric loop with integrated SFS junction by the appearance of spontaneous supercurrents\cite{bauer2004}. Nowadays, the $\pi$-junction state of SFS attracts much attention due to applications in the flux-quantum logic based memory cells \cite{terzioglu1998,ustinov2003,khabipov2010,bakurskiy2018protected} and superconducting qubit implementations \cite{yamashita2005,feofanov2010}. 

  \begin{figure*}[htb!]
\includegraphics[width=0.2\linewidth,angle=-90,trim={0.5cm 0.5cm 0.5cm 0.5cm},clip]{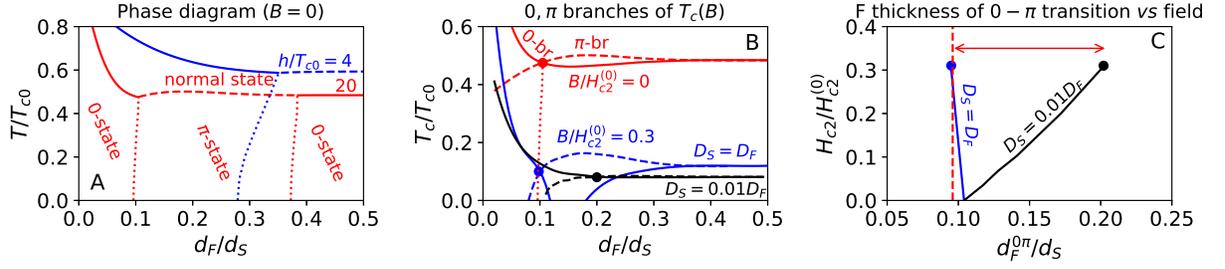}
 \caption{\label{f:Tc2} 
  (Color online) (A) Zero-field $B=0$ phase diagram for the ground states of SFS junction with red and blue curves corresponding to $h/T_{c0}=20$ and $h/T_{c0}=4$, respectively. Boundaries
  restrict the 0 state either from normal one at higher temperatures (solid) or $\pi$-state at different $d_F$ (dotted). The transition from $\pi$ to normal state is shown by dashed curve. For panel A we have considered $D_S=D_F$. (B) Field-dependent critical temperatures of the transitions to $0$ and $\pi$ states shown by solid and dashed curves, respectively, and their first crossing points (dots). Red: $B/H_{c2}^{(0)}=0$; blue, black: $0.3$ for $D_S/D_F=1$ and $0.01$, respectively, and $H_{c2}^{(0)}$ is the upper critical field in the bulk. 
 Vertical dotted line is 0-$\pi$ transition for $B=0$.  %
  (C) Field dependencies of  the 0-$\pi$ transition thickness $d_F^{0\pi}$ determined by the first crossing of the 0 and $\pi$ branches of $T_c(B)$ for different $D_S/D_F$ ratios. Note that each point on the curves corresponds to the temperature $T_c(B=H_{c2},d_F^{0\pi})$. Dots indicate the case $T_c(B=H_{c2},d_F^{0\pi})\to 0$. For comparison the value of
   $d_F^{0\pi} (B=0,T=0)$ is shown by dashed line. The maximal interval of the field-controlled 0-$\pi$ transition occuring at $T\to 0$ is shown by the red arrow. 
  In (B, C) $h/T_{c0}= 20$, 
  and in all panels
  $d_S=3.3\xi$, where 
   $2\pi T_{c0}\xi^2=D_F$.
    }
  \end{figure*}
  

Although the 0-$\pi$ transition can be observed by changing  the temperature for weak ferromagnets with small $h$ \cite{ryazanov2001}, this can be more challenging in systems with  $h\gg T_{c0}$ where we denote $T_{c0}$ to be the bulk critical temperature of the superconducting layer. 
This can be illustrated using the temperature-thickness phase diagram in Fig. \ref{f:Tc2}A calculated at $B=0$  as described below for the structure shown 
schematically in upper left panel in Fig. \ref{f:sfs}. 
For large exchange fields, $h=20 T_{c0}$,  the boundary between $0$ and $\pi$
states is almost vertical, so that one should control $d_F$ with 
very high precision to spot the region of temperature-controlled transition. 
The origin of this behaviour can be understood by considering the complex length $\xi_F^{-1} = \sqrt{ (T + ih )/D_F}$ which determines the behaviour of
superconducting correlations in the F layer characterized by diffusion coefficient $D_F$. For $h\gg  T_{c0}$ the scale is temperature-independent for the considered regime $T<T_{c0}$ and hence the $0$-$\pi$ switching occurs at the same $d_F$ for all temperatures. 
Thus the only 
way to switch SFS regularly from 0 to $\pi$ state in this case
is to scan over $d_F$ which requires fabrication and measuring many samples. 

Here we show that this situation can be improved by introducing the additional control parameter which is the magnetic field $B$ 
perpendicular to the layers. 
%
%
Unlike temperature-driven 0-$\pi$ transition which requires weak F and fine tuning of the F thickness, we show that the interval of F thicknesses suitable for field-driven transition can be made arbitrarily wide for any exchange field by reducing the diffusion constant of S with respect to the one in F.

In the case of applied perpendicular magnetic field, the scale of oscillations in the F layer is determined at small  temperatures $T\ll q$ by 
$\xi_{F}^{-1}(B) = \sqrt{ (q +ih )/D_F}$ where $q=eBD_F $. 
Then, even for arbitrary large $h$ the orbital effect can introduce significant shift of $\xi_F$ if $q \sim h$. This conditions can be achieved if the orbital depairing can be made sufficiently strong. The largest values of $q$ can be obtained near the upper critical field $B=H_{c2}$.  By taking into account estimation $eH_{c2}D_S\sim1$, 
 the regime $q\sim h$ requires the diffusion coefficient in the superconductor $D_S$ much smaller than that in the ferromagnet $D_S\ll D_F$.
This condition can be always achieved by intentionally adding impurities and decreasing electron scattering time in dirty S. The effect is demonstrated in Figs. \ref{f:Tc2}B,D as significant shift of the F thickness segregating 0 and $\pi$ states from its zero-field value by applying magnetic field.     




 In the intermediate region of perpendicular 
magnetic fields $0<B<H_{c2}$ the SFS junction is in the mixed state, which means that it is 
pierced by the Abrikosov vortex lines. The natural question is how the $0$ and $\pi$ superconducting states manifest themselves in the vortex behaviour. In principle, the discrepancy between distributions of the gap order parameter in these states results in the different response to the applied magnetic field. This was revealed recently, for instance, by superfluid-density measurements \cite{hinton2016}.

The characteristic feature of the mixed state is a non-zero resistivity which occurs due to the dissipative motion of mobile vortex
lines in the superconducting environment.  Below we show that distinct gap profiles of the 0 and $\pi$-state lead also to the difference in flux-flow resistivity of SFS. We demonstrate that one can detect $0$-$\pi$ transitions measuring the 
qualitative change in the dependence of resisitivity on $d_F$ 
in the perpendicular magnetic field. 


Below we present theoretical description  consistent with available flux-flow resistivity data for SFS \cite{Pompeo} and discuss low-field flux-flow resistivity experiment, where resisitivity of SFS is proportional to the vortex density in agreement with Bardeen-Stephen theory. We argue that in the $\pi$-state the relevant numeric proportionality coefficient exhibits universal $h$-independent behaviour providing a way to distinguish between the 0 and $\pi$ state of SFS by single flux-flow resisitivity measurement.

{\bf Model.} We start with the formalisim of quasiclassical Green's function (GF) \cite{Schmid1975} generalized to describe non-equilibrium spin states in diffusive superconductors \cite{bergeret2018}, $\check{g}(t_1,t_2,\bf r) = \left(%
 \begin{array}{cc}
  \hat g^R &  \hat g^K \\
  0 &  \hat g^A \\
 \end{array}\label{eq:GF0}
 \right)$, where $\hat g^{R/A/K}$ are the retarded/advanced/Keldysh components which are determined by the  Keldysh-Usadel equation 
  \begin{equation}  \label{Eq:UsadelGen}  
  \{\hat\tau_3\partial_t, \check g \}_t = 
 \hat\partial_{\bm r} \hat D  ( \check g
\circ \hat\partial_{\bm r} \check g) 
 - i [\hat\tau_3\hat H , \check g ]_t ,
   \end{equation}
where $\hat D$ is the diffusivity tensor which can be anisotropic and space-dependent,  
$\hat H=  \hat{\bm\sigma}\bm h-\hat\tau_1\hat\Delta $,  and $\hat\tau_i$ and $\hat\sigma_i$ ($i=0,1,2,3$) are Pauli matrices in Nambu and spin space, $\bm h$ is the exchange field.
The gap function $\hat\Delta=|\Delta|e^{-i\hat\tau_3\varphi}$, where $\varphi$ is the gap phase, is determined by the self-consistency condition
\begin{align} \label{Eq:SelfConsistency}
\qquad \Delta= \pi\lambda {\rm Tr}\hat g^K_{12} (t,t)/4,
\end{align}
where $\lambda$ is the coupling constant finite in S layers. 
In Eq. (\ref{Eq:UsadelGen}), the commutator operator is defined as
   $[X, g]_t= X(t_1) g(t_1,t_2)- g(t_1,t_2) X(t_2)$, similarly for anticommutator 
   $\{,\}_t$. The symbolic product operator is given by
   $ (A\circ B) (t_1,t_2) = \int dt A(t_1,t)B(t,t_2)$ and covariant differential superoperator reads as
$\hat \partial_{\bm r}  = 
\partial_{\bm r} - ie\bm A [\hat\tau_3 , \cdot ]$,
%
where $e$ is the elementary charge. 
The diffusion coefficient is different in S and F regions. For simplicity we assume that F is isotropic $D_z=D_{x,y} = D_F $ while S is anisotropic with $ D_z = D_F$ and $ D_{x,y} = D_S$. The anisotropy assumption does not affect results qualitatively since they rely on the difference of diffusion coefficients in 
the direction perpendicular to the magnetic field, that is along the layers. 

First, we start with the  equilibrium problem of the magnetic-field driven $0$-$\pi$ transitions. Our goal is to find the range of parameters where the system undergoes this transition with changing magnetic field from $0$ to $H_{c2}$ in the direction perpendicular to SF interface, $\bm B=B\bm z$.
To determine such parameters it is enough to compare the states at the end points of this interval, namely at  $B=0$ and at $B=H_{c2}$.
We assume $\bm h=h\bm z$ in F layer, put $\hat g (t_1,t_2) = \int_{-\infty}^{\infty} \hat g(\varepsilon, t) e^{-i\varepsilon
(t_1-t_2) } \frac{d\varepsilon}{2\pi} $, where $t=(t_1+t_2)/2$, and use gradient expansion for time-convolution products to obtain 
equations for the temperature GF by replacing $-i\varepsilon $ with Matsubara frequency $\omega_n$.

 In the absence of magnetic field $B=0$, we
  determined the lowest-energy state 
  of the SFS system on the $T$, $d_F$ plane by evaluating the free energy \cite{1909.00992} using self-consistent distribution of $\Delta (z)$ and  corresponding GF\cite{SM}.
  %
 By comparing numerical values of the 0 and $\pi$ branches of free energy, we obtained the first-order 0-$\pi$ transition lines shown in Fig. \ref{f:Tc2}A. Previosly, thermodynamic 0-$\pi$ transition was discussed only within Ginzburg-Landau theory \cite{samokhvalov2015}. 

One can see that for  parameters $h\gg T_{c0}$ typical for strong ferromagnets such as Co\cite{} the
0-$\pi$ transition curve is almost vertical, that is the threshold thickness $d_F^{0\pi}$ depends on the temperature very weakly. Practically, this means that it quite difficult to 
choose $d_F$ in the range where SFS system has temperature-controlled $0$-$\pi$ transition. 
 
 To find how magnetic field changes critical temperatures of the $0$ and $\pi$-states we generalize  the multi-mode approach used previously for SF bilayers
\cite{Fominov2002, Krunavakarn2006}.
Using the symmetry of solutions we reduce the SFS problem to that of the SF bilayer with different 
boundary conditions at free F interface corresponding to $0$ and $\pi$ states. 
 We consider gauge ${\bm A}=\bm yBx$ and apply the Abrikosov ansatz 
\begin{align}\label{Abrikosov}
&\Delta=\sum_m C_m e^{impy} \tilde\Delta(x-mx_0,z),
\\
&\hat g_{12n}=\sum_{m,\sigma} C_m e^{impy}
f_{\sigma n}(x-mx_0,z)\hat\sigma_\sigma.\label{f}
\end{align}
Here anomalous Matsubara GF (\ref{f}) is extended into spin space by introducing $2\hat \sigma_\sigma=\hat\sigma_0+\sigma\hat \sigma_3$, where $\sigma=\pm$. Other notations are conventional for lattice solution, namely, $|C_m|=1$, $p$ is defined by lattice symmetry, $x_0=pL_H^2$ and $L_H^{-2}=2eB$. Next we separate variables 
$f_{\sigma n}(x,z)=\Psi_0(x)\alpha_{\sigma n}(z)$ and $\tilde \Delta(x,z)=\Psi_0(x)\beta(z)$, where $\Psi_0=e^{-L_H^{-2}x^2/2}$ is zero Landau level eigenfunction, to obtain Usadel Eq. in the form
\begin{align}\label{ddd}
&D_z\partial_z^2\alpha_{\sigma n}- 2[(\omega_n+q + i \sigma h)\alpha_{\sigma n}+i\beta ]=0,
\end{align} 
together with self-consistency condition $\beta=\lambda\pi i T\sum_{\sigma,n\geq 0} \alpha_{\sigma n}$. Here $q=eBD_x$ is the 
orbital energy.
We solve\cite{SM} Eq. (\ref{ddd}) together with boundary conditions and self-consistency equation by means of multi-mode approach \cite{Fominov2002, Krunavakarn2006} yielding the the upper critical field $H_{c2} = H_{c2}(T)$ or field-dependent critical temperature $T_c=T_c(B)$.

 The resulting dependencies $T_c=T_c(d_F)$
for different $B$  are shown in  Fig. \ref{f:Tc2}B. 
For small magnetic fields, 
we have intersecting $0$ and $\pi$ branches resulting in the oscillatory behaviour of $T_c(d_F)$\cite{Fominov2002,jiang1995,samokhvalov2017}. For larger $B$, there appear intervals of $d_F$
with only one stable state, either $0$ or $\pi$
as shown by blue curves in Fig. \ref{f:Tc2}B. 
The  $0$-$\pi$ transitions occur at the values of thickness $d_F^{0\pi}$ determined by the intersection of 0 and $\pi$ branches of $T_c(d_F)$ (shown by dots in Fig. \ref{f:Tc2}B). 
Fig. \ref{f:Tc2}C demonstrates the magnetic field dependence of $d_F^{0\pi}$ confirming our qualitative arguments about its high sensitivity to the ratio of diffusion coefficients in F and S layers.
For $D_F/D_S> h/T_{c0}$ (black line in Fig.\ref{f:Tc2}C ) there is a strong variation of threshold thickness with field as compared with almost no dependence of $d_F^{0\pi}$ in the opposite case (blue line in Fig.\ref{f:Tc2}C). 

To understand the SFS behaviour under applied magnetic field it is enough to compare the 
endpoints which are the states  at $B=0$ and at  $B=H_{c2}$ shown in Figs. \ref{f:Tc2}A,C, respectively.
In Figs. \ref{f:Tc2}C the $0$ states at $B=H_{c2}$ are on  the left of the corresponding solid curve while $\pi$ states at $B=0$ and $T=0$ are on the right of the dashed line. For larger $T$ the  shift of dashed line is negligible as can be inferred from Fig.\ref{f:Tc2}A. 
From comparison of dashed and solid black lines in Figs. \ref{f:Tc2}C one can see that for $D_S=0.01 D_F$
there is a wide interval of $d_F$ where the  $0$-$\pi$ transition {\it with necessity} occurs when varying magnetic field from $0$ to $H_{c2}$ at fixed $T$ and $d_F$. This interval bounded by $d_F^{0\pi}(H_{c2})$ curve and  $d_F^{0\pi}(B=0)$ value is shown by the red arrow in Fig. \ref{f:Tc2}C.

 \begin{figure}[htb!]
\includegraphics[width=0.95\linewidth,trim={0.5cm 0.5cm 0.5cm 0cm},clip]{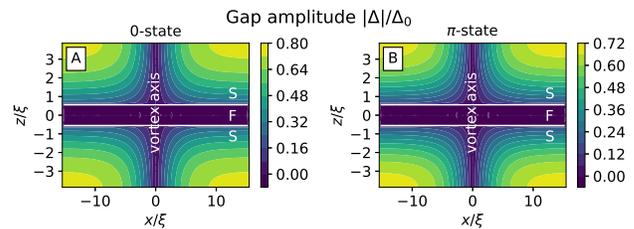}
 \caption{\label{f:GapProfiles} 
  (Color online) 
  Gap profile $|\Delta|/\Delta_0$ normalized to the bulk gap $\Delta_0$ in the vortex cell of SFS for the 0 (A) and $\pi$-state (B). Calculations have been done for $d_F/d_S=0.17$, $h/T_{c0}=6$ and $D_S=D_F$. White horizontal lines correspond to SF interfaces. 
  }
  \end{figure}

 {\bf Flux-flow resistivity}.
 At intermediate values of magnetic field $0<B<H_{c2}$ 
 SFS system is in the mixed state consisting of Abrikosov vortex (AV) lines shown schematically in lower panels of Fig. \ref{f:sfs}.
 Vortex structure transforms due to the proximity effect\cite{golubov2016abrikosov,stolyarov2018expansion}.
 Such a transformation is different in the $0$ and $\pi$ states of the SFS system which affects their dynamical properties as shown below. 
 
 \begin{figure*}[htb!]
\includegraphics[width=0.19\linewidth,angle=-90,trim={0.5cm 0.5cm 0.5cm 0.5cm},clip]{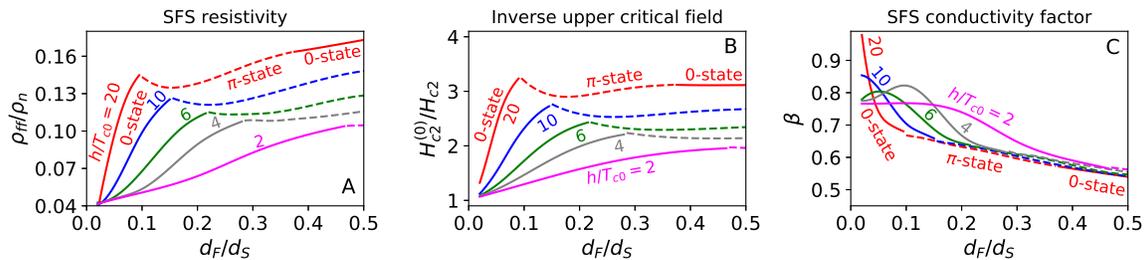}
 \caption{\label{f4}
   (A) The crossover of 
  $\rho_{ff}$ at the
  $0$-$\pi$ transition. 
  (B) Inverse upper critical field $H_{c2}^{-1}(d_F)$.
  %
  (C) The numerical factor $\beta$ in Bardeen-Stephen expression. 
    In (A-C)  $T=0.055T_{c0}\ll T_c$, $D_S=D_F$ and colors indicate different $h/T_{c0} = 2, 4, 6, 10, 20$. Solid and dashed curves are stable 0 and $\pi$ branches. 
In (A), (C)  $B=0.03H_{c2}^{(0)}\ll H_{c2}$.
   }
  \end{figure*}
 
To calculate the structure of individual  vortices at finite magnetic fields we use the circular cell approximation
 \cite{ihle1971b,WattsTobin1974,Rammer1987,Rammer1988}, where the unit cell of the hexagonal vortex lattice hosting a single vortex is replaced by a circular cell with the centre at the point of superconducting phase singularity. Inside circular cell, the gap and magnetic field distributions are taken radially symmetric with respect to the cell centre. 
 At that, the circular-cell radius is uniquely defined by magnetic induction, $r_c=\sqrt{\phi_0/(\pi B)}$ so that there is exactly one flux quantum $\phi_0=\pi/e$ passing through the unit vortex cell\cite{SM}. Calculated gap profile inside the cell is shown in Fig. \ref{f:GapProfiles}. 

The controlled motion of the vortices can be produced by applying transport current $\bm j_{tr}$ which exerts the Lorentz force $\bm F_L=\phi_0{\bm j}_{tr}\times {\bm z}$ on each vortex due to interaction with its local magnetic field. Vortex motion with velocity $\bm v_L$ produces perpendicular electric field ${\bm E}={\bm B}\times {\bm v}_L$ as shown in Fig. \ref{f:sfs}. This field causes energy dissipation due to the ohmic losses inside the normal vortex core which can be expressed as the viscous friction ${\bm F}=-\eta {\bm v}_L$, where $\eta$ is vortex viscosity. In the steady-state regime, $\bm F+\bm F_L=0$, we obtain ${\bm E}=\rho_{ff}{\bm j}_{tr}$, where $\rho_{ff}=\phi_0 B/\eta$ is flux-flow resistivity. 
 
 To calculate viscosity $\eta$, we consider microscopic expression\cite{LO,kopninbook}  for the force $\bm F$ acting due to non-equilibrium environment\cite{SM}. The latter is determined by the vortex-motion induced deviations of electron distribution function from the Fermi-Dirac one which obey kinetic equations derived \cite{SM} from the Keldysh part of the Keldysh-Usadel Eq. (\ref{Eq:UsadelGen}). The coefficients
 in kinetic equations are determined by the vortex structure that we find from equilibrium problem as explained above.  
We consider low-temperature regime where the nonequlibrium states have subgap energies and therefore their contributions relax at the distances of the order of coherence length. This is different from the vicinity of $T_c$ where vortex motion in multilayered systems is determined by the renormalization of long-range charge imbalance mode\cite{mel1996inertial}.

In Fig. \ref{f4}B we show the calculated $\rho_{ff}$ for stable parts of the 0 and $\pi$  branches. 
The intersection of these branches points to the first-order 0-$\pi$ transition whose position scales with $\xi_F\sim1/\sqrt{h}$. 
The crossover behaviour of $\rho_{ff}$ at $0$-$\pi$ transition can be understood qualitatively using the Bardeen-Stephen expression $\rho_{ff}/\rho_{n}=\beta^{-1} B/H_{c2}$, where  $\beta\sim 1$ is determined by the particular microscopic model and $\rho_n$ is normal-state resistivity. 
The inverse upper critical field of SFS trilayer\cite{SM,radovic1991,Krunavakarn2006} $H^{-1}_{c2} (d_F)$ is shown in Fig. \ref{f4}B. 
 %
The variations of $H^{-1}_{c2} (d_F)$ follow closely the behaviour of $\rho_{ff}$. Their oscillation in the vicinity of the 0-$\pi$ transition is caused by the superconductivity suppression with $d_F$ in the 0 and its enhancement in the $\pi$ state.
In the considered low-temperature regime  $\beta\approx 0.77$ for usual single-band dirty superconductors\cite{vs}. Fig. \ref{f4}D demonstrates dependencies $\beta(d_F)$ for SFS sandwich. We see that bulk value is approached in the limit $d_F\to0$, that is in the absence of F layer. For finite $d_F$, the function $\beta(d_F)$ passes in the 0-state through the maximum whose height exceeds bulk value 0.77 for not very weak F. At that, in the $\pi$-state $\beta<0.77$  approaches universal $h$-independent asymptotic weakly varying with $d_F$. These signatures of $\beta$ can be used for distinguishing the state of SFS with the help of the single flux-flow resistivity experiment without fabricating and measuring many samples. 

Results shown in Fig. \ref{f4}A are in qualitative agreement with  measurements demonstrating the increase followed by the saturation of flux-flow resistivity in SFS trilayer with the growth of $d_F$ \cite{Pompeo}. Although oscillations of $\rho_{ff}(d_F)$ and $0$-$\pi$ transition point were not directly detected in this experiment, even in such a case flux-flow resisitivity measurements allow to distinguish between samples in the 0 and $\pi$ states by means of the $\beta$ value as discussed above.

To conclude, we have demonstrated the possibility of the $0$-$\pi$ transitions in the SFS structure driven by the perpendicular magnetic field. These transitions can be achieved in the wide interval of the F layer thicknesses provided the S layer has much smaller diffusion coefficient than F layer. In contrast to the temperature-driven ones, the  magnetic field-driven $0$-$\pi$ transitions can be realized in principle for arbitrary large exchange field $h\gg T_{c}$. Besides that we have found indications of $0$-$\pi$ transitions in the flux-flow conductivity of SFS structure. This behaviour is in the qualitative agreement with experimental observations.  

This work was supported by the Academy of Finland
(Project No. 297439), Russian Science Foundation (Grant No. 19-19-00594) and the European Regional Development Fund (Mobilitas Pluss grant MOBTP152). It is our pleasure to acknowledge discussions with M. Yu. Kupriyanov, M. M. Khapaev, N. Pompeo 
and A.S. Mel'nikov.

\bibliography{sfs}

\begin{thebibliography}{37}%
\makeatletter
\providecommand \@ifxundefined [1]{%
 \@ifx{#1\undefined}
}%
\providecommand \@ifnum [1]{%
 \ifnum #1\expandafter \@firstoftwo
 \else \expandafter \@secondoftwo
 \fi
}%
\providecommand \@ifx [1]{%
 \ifx #1\expandafter \@firstoftwo
 \else \expandafter \@secondoftwo
 \fi
}%
\providecommand \natexlab [1]{#1}%
\providecommand \enquote  [1]{``#1''}%
\providecommand \bibnamefont  [1]{#1}%
\providecommand \bibfnamefont [1]{#1}%
\providecommand \citenamefont [1]{#1}%
\providecommand \href@noop [0]{\@secondoftwo}%
\providecommand \href [0]{\begingroup \@sanitize@url \@href}%
\providecommand \@href[1]{\@@startlink{#1}\@@href}%
\providecommand \@@href[1]{\endgroup#1\@@endlink}%
\providecommand \@sanitize@url [0]{\catcode `\\12\catcode `\$12\catcode
  `\&12\catcode `\#12\catcode `\^12\catcode `\_12\catcode `\%12\relax}%
\providecommand \@@startlink[1]{}%
\providecommand \@@endlink[0]{}%
\providecommand \url  [0]{\begingroup\@sanitize@url \@url }%
\providecommand \@url [1]{\endgroup\@href {#1}{\urlprefix }}%
\providecommand \urlprefix  [0]{URL }%
\providecommand \Eprint [0]{\href }%
\providecommand \doibase [0]{http://dx.doi.org/}%
\providecommand \selectlanguage [0]{\@gobble}%
\providecommand \bibinfo  [0]{\@secondoftwo}%
\providecommand \bibfield  [0]{\@secondoftwo}%
\providecommand \translation [1]{[#1]}%
\providecommand \BibitemOpen [0]{}%
\providecommand \bibitemStop [0]{}%
\providecommand \bibitemNoStop [0]{.\EOS\space}%
\providecommand \EOS [0]{\spacefactor3000\relax}%
\providecommand \BibitemShut  [1]{\csname bibitem#1\endcsname}%
\let\auto@bib@innerbib\@empty
\bibitem [{\citenamefont {Buzdin}\ \emph {et~al.}(1982)\citenamefont {Buzdin},
  \citenamefont {Bulaevskii},\ and\ \citenamefont {Panyukov}}]{buzdin1982}%
  \BibitemOpen
  \bibfield  {author} {\bibinfo {author} {\bibfnamefont {A.~I.}\ \bibnamefont
  {Buzdin}}, \bibinfo {author} {\bibfnamefont {L.}~\bibnamefont {Bulaevskii}},
  \ and\ \bibinfo {author} {\bibfnamefont {S.}~\bibnamefont {Panyukov}},\
  }\href@noop {} {\bibfield  {journal} {\bibinfo  {journal} {JETP. Lett.}\
  }\textbf {\bibinfo {volume} {35}},\ \bibinfo {pages} {147} (\bibinfo {year}
  {1982})}\BibitemShut {NoStop}%
\bibitem [{\citenamefont {Buzdin}\ and\ \citenamefont
  {Kupriyanov}(1991)}]{buzdin1991}%
  \BibitemOpen
  \bibfield  {author} {\bibinfo {author} {\bibfnamefont {A.~I.}\ \bibnamefont
  {Buzdin}}\ and\ \bibinfo {author} {\bibfnamefont {M.~Y.}\ \bibnamefont
  {Kupriyanov}},\ }\href@noop {} {\bibfield  {journal} {\bibinfo  {journal}
  {JETP. Lett.}\ }\textbf {\bibinfo {volume} {53}},\ \bibinfo {pages} {321}
  (\bibinfo {year} {1991})}\BibitemShut {NoStop}%
\bibitem [{\citenamefont {Buzdin}(2005)}]{Buzdin2005}%
  \BibitemOpen
  \bibfield  {author} {\bibinfo {author} {\bibfnamefont {A.~I.}\ \bibnamefont
  {Buzdin}},\ }\href@noop {} {\bibfield  {journal} {\bibinfo  {journal} {Rev.
  Mod. Phys.}\ }\textbf {\bibinfo {volume} {77}},\ \bibinfo {pages} {935}
  (\bibinfo {year} {2005})}\BibitemShut {NoStop}%
\bibitem [{\citenamefont {Lyuksyutov}\ and\ \citenamefont
  {Pokrovsky}(2005)}]{lyuksyutov2005}%
  \BibitemOpen
  \bibfield  {author} {\bibinfo {author} {\bibfnamefont {I.~F.}\ \bibnamefont
  {Lyuksyutov}}\ and\ \bibinfo {author} {\bibfnamefont {V.~L.}\ \bibnamefont
  {Pokrovsky}},\ }\href@noop {} {\bibfield  {journal} {\bibinfo  {journal}
  {Adv. Phys.}\ }\textbf {\bibinfo {volume} {54}},\ \bibinfo {pages} {67}
  (\bibinfo {year} {2005})}\BibitemShut {NoStop}%
\bibitem [{\citenamefont {Ryazanov}\ \emph {et~al.}(2001)\citenamefont
  {Ryazanov}, \citenamefont {Oboznov}, \citenamefont {Rusanov}, \citenamefont
  {Veretennikov}, \citenamefont {Golubov},\ and\ \citenamefont
  {Aarts}}]{ryazanov2001}%
  \BibitemOpen
  \bibfield  {author} {\bibinfo {author} {\bibfnamefont {V.~V.}\ \bibnamefont
  {Ryazanov}}, \bibinfo {author} {\bibfnamefont {V.~A.}\ \bibnamefont
  {Oboznov}}, \bibinfo {author} {\bibfnamefont {A.~Y.}\ \bibnamefont
  {Rusanov}}, \bibinfo {author} {\bibfnamefont {A.~V.}\ \bibnamefont
  {Veretennikov}}, \bibinfo {author} {\bibfnamefont {A.~A.}\ \bibnamefont
  {Golubov}}, \ and\ \bibinfo {author} {\bibfnamefont {J.}~\bibnamefont
  {Aarts}},\ }\href@noop {} {\bibfield  {journal} {\bibinfo  {journal} {Phys.
  Rev. Lett.}\ }\textbf {\bibinfo {volume} {86}},\ \bibinfo {pages} {2427}
  (\bibinfo {year} {2001})}\BibitemShut {NoStop}%
\bibitem [{\citenamefont {Kontos}\ \emph {et~al.}(2002)\citenamefont {Kontos},
  \citenamefont {Aprili}, \citenamefont {Lesueur}, \citenamefont {Gent},
  \citenamefont {Stephanidis},\ and\ \citenamefont {Boursier}}]{kontos2002}%
  \BibitemOpen
  \bibfield  {author} {\bibinfo {author} {\bibfnamefont {T.}~\bibnamefont
  {Kontos}}, \bibinfo {author} {\bibfnamefont {M.}~\bibnamefont {Aprili}},
  \bibinfo {author} {\bibfnamefont {J.}~\bibnamefont {Lesueur}}, \bibinfo
  {author} {\bibfnamefont {F.}~\bibnamefont {Gent}}, \bibinfo {author}
  {\bibfnamefont {B.}~\bibnamefont {Stephanidis}}, \ and\ \bibinfo {author}
  {\bibfnamefont {R.}~\bibnamefont {Boursier}},\ }\href@noop {} {\bibfield
  {journal} {\bibinfo  {journal} {Phys. Rev. Lett.}\ }\textbf {\bibinfo
  {volume} {89}},\ \bibinfo {pages} {137007} (\bibinfo {year}
  {2002})}\BibitemShut {NoStop}%
\bibitem [{\citenamefont {Shelukhin}\ \emph {et~al.}(2006)\citenamefont
  {Shelukhin}, \citenamefont {Tsukernik}, \citenamefont {Karpovski},
  \citenamefont {Blum}, \citenamefont {Efetov}, \citenamefont {Volkov},
  \citenamefont {Champel}, \citenamefont {Eschrig}, \citenamefont
  {L{\"o}fwander}, \citenamefont {Schoen} \emph
  {et~al.}}]{shelukhin2006observation}%
  \BibitemOpen
  \bibfield  {author} {\bibinfo {author} {\bibfnamefont {V.}~\bibnamefont
  {Shelukhin}}, \bibinfo {author} {\bibfnamefont {A.}~\bibnamefont
  {Tsukernik}}, \bibinfo {author} {\bibfnamefont {M.}~\bibnamefont
  {Karpovski}}, \bibinfo {author} {\bibfnamefont {Y.}~\bibnamefont {Blum}},
  \bibinfo {author} {\bibfnamefont {K.}~\bibnamefont {Efetov}}, \bibinfo
  {author} {\bibfnamefont {A.}~\bibnamefont {Volkov}}, \bibinfo {author}
  {\bibfnamefont {T.}~\bibnamefont {Champel}}, \bibinfo {author} {\bibfnamefont
  {M.}~\bibnamefont {Eschrig}}, \bibinfo {author} {\bibfnamefont
  {T.}~\bibnamefont {L{\"o}fwander}}, \bibinfo {author} {\bibfnamefont
  {G.}~\bibnamefont {Schoen}},  \emph {et~al.},\ }\href@noop {} {\bibfield
  {journal} {\bibinfo  {journal} {Physical Review B}\ }\textbf {\bibinfo
  {volume} {73}},\ \bibinfo {pages} {174506} (\bibinfo {year}
  {2006})}\BibitemShut {NoStop}%
\bibitem [{\citenamefont {Oboznov}\ \emph {et~al.}(2006)\citenamefont
  {Oboznov}, \citenamefont {Bolginov}, \citenamefont {Feofanov}, \citenamefont
  {Ryazanov},\ and\ \citenamefont {Buzdin}}]{oboznov2006}%
  \BibitemOpen
  \bibfield  {author} {\bibinfo {author} {\bibfnamefont {V.~A.}\ \bibnamefont
  {Oboznov}}, \bibinfo {author} {\bibfnamefont {V.~V.}\ \bibnamefont
  {Bolginov}}, \bibinfo {author} {\bibfnamefont {A.~K.}\ \bibnamefont
  {Feofanov}}, \bibinfo {author} {\bibfnamefont {V.~V.}\ \bibnamefont
  {Ryazanov}}, \ and\ \bibinfo {author} {\bibfnamefont {A.~I.}\ \bibnamefont
  {Buzdin}},\ }\href@noop {} {\bibfield  {journal} {\bibinfo  {journal} {Phys.
  Rev. Lett.}\ }\textbf {\bibinfo {volume} {96}},\ \bibinfo {pages} {197003}
  (\bibinfo {year} {2006})}\BibitemShut {NoStop}%
\bibitem [{\citenamefont {Bauer}\ \emph {et~al.}(2004)\citenamefont {Bauer},
  \citenamefont {Bentner}, \citenamefont {Aprili}, \citenamefont {Rocca},
  \citenamefont {Reinwald}, \citenamefont {Wegscheider},\ and\ \citenamefont
  {Strunk}}]{bauer2004}%
  \BibitemOpen
  \bibfield  {author} {\bibinfo {author} {\bibfnamefont {A.}~\bibnamefont
  {Bauer}}, \bibinfo {author} {\bibfnamefont {J.}~\bibnamefont {Bentner}},
  \bibinfo {author} {\bibfnamefont {M.}~\bibnamefont {Aprili}}, \bibinfo
  {author} {\bibfnamefont {M.~L.~D.}\ \bibnamefont {Rocca}}, \bibinfo {author}
  {\bibfnamefont {M.}~\bibnamefont {Reinwald}}, \bibinfo {author}
  {\bibfnamefont {W.}~\bibnamefont {Wegscheider}}, \ and\ \bibinfo {author}
  {\bibfnamefont {C.}~\bibnamefont {Strunk}},\ }\href@noop {} {\bibfield
  {journal} {\bibinfo  {journal} {Phys. Rev. Lett.}\ }\textbf {\bibinfo
  {volume} {92}},\ \bibinfo {pages} {217001} (\bibinfo {year}
  {2004})}\BibitemShut {NoStop}%
\bibitem [{\citenamefont {Terzioglu}\ and\ \citenamefont
  {Beasley}(1998)}]{terzioglu1998}%
  \BibitemOpen
  \bibfield  {author} {\bibinfo {author} {\bibfnamefont {E.}~\bibnamefont
  {Terzioglu}}\ and\ \bibinfo {author} {\bibfnamefont {M.~R.}\ \bibnamefont
  {Beasley}},\ }\href@noop {} {\bibfield  {journal} {\bibinfo  {journal} {IEEE
  Trans. Appl. Supercon.}\ }\textbf {\bibinfo {volume} {8}},\ \bibinfo {pages}
  {48} (\bibinfo {year} {1998})}\BibitemShut {NoStop}%
\bibitem [{\citenamefont {Ustinov}\ and\ \citenamefont
  {Kaplunenko}(2003)}]{ustinov2003}%
  \BibitemOpen
  \bibfield  {author} {\bibinfo {author} {\bibfnamefont {A.~V.}\ \bibnamefont
  {Ustinov}}\ and\ \bibinfo {author} {\bibfnamefont {V.~K.}\ \bibnamefont
  {Kaplunenko}},\ }\href@noop {} {\bibfield  {journal} {\bibinfo  {journal} {J.
  Appl. Phys.}\ }\textbf {\bibinfo {volume} {94}},\ \bibinfo {pages} {5405}
  (\bibinfo {year} {2003})}\BibitemShut {NoStop}%
\bibitem [{\citenamefont {Khabipov}\ \emph {et~al.}(2010)\citenamefont
  {Khabipov}, \citenamefont {Balashov}, \citenamefont {Maibaum}, \citenamefont
  {Zorin}, \citenamefont {Oboznov}, \citenamefont {Bolginov}, \citenamefont
  {Rossolenko},\ and\ \citenamefont {Ryazanov}}]{khabipov2010}%
  \BibitemOpen
  \bibfield  {author} {\bibinfo {author} {\bibfnamefont {M.~I.}\ \bibnamefont
  {Khabipov}}, \bibinfo {author} {\bibfnamefont {D.~V.}\ \bibnamefont
  {Balashov}}, \bibinfo {author} {\bibfnamefont {F.}~\bibnamefont {Maibaum}},
  \bibinfo {author} {\bibfnamefont {A.~B.}\ \bibnamefont {Zorin}}, \bibinfo
  {author} {\bibfnamefont {V.~A.}\ \bibnamefont {Oboznov}}, \bibinfo {author}
  {\bibfnamefont {V.~V.}\ \bibnamefont {Bolginov}}, \bibinfo {author}
  {\bibfnamefont {A.~N.}\ \bibnamefont {Rossolenko}}, \ and\ \bibinfo {author}
  {\bibfnamefont {V.~V.}\ \bibnamefont {Ryazanov}},\ }\href@noop {} {\bibfield
  {journal} {\bibinfo  {journal} {Supercond. Sci. Technol.}\ }\textbf {\bibinfo
  {volume} {23}},\ \bibinfo {pages} {045032} (\bibinfo {year}
  {2010})}\BibitemShut {NoStop}%
\bibitem [{\citenamefont {Bakurskiy}\ \emph {et~al.}(2018)\citenamefont
  {Bakurskiy}, \citenamefont {Klenov}, \citenamefont {Soloviev}, \citenamefont
  {Pugach}, \citenamefont {Kupriyanov},\ and\ \citenamefont
  {Golubov}}]{bakurskiy2018protected}%
  \BibitemOpen
  \bibfield  {author} {\bibinfo {author} {\bibfnamefont {S.}~\bibnamefont
  {Bakurskiy}}, \bibinfo {author} {\bibfnamefont {N.}~\bibnamefont {Klenov}},
  \bibinfo {author} {\bibfnamefont {I.}~\bibnamefont {Soloviev}}, \bibinfo
  {author} {\bibfnamefont {N.}~\bibnamefont {Pugach}}, \bibinfo {author}
  {\bibfnamefont {M.~Y.}\ \bibnamefont {Kupriyanov}}, \ and\ \bibinfo {author}
  {\bibfnamefont {A.}~\bibnamefont {Golubov}},\ }\href@noop {} {\bibfield
  {journal} {\bibinfo  {journal} {Applied physics letters}\ }\textbf {\bibinfo
  {volume} {113}},\ \bibinfo {pages} {082602} (\bibinfo {year}
  {2018})}\BibitemShut {NoStop}%
\bibitem [{\citenamefont {Yamashita}\ \emph {et~al.}(2005)\citenamefont
  {Yamashita}, \citenamefont {Tanikawa}, \citenamefont {Takahashi},\ and\
  \citenamefont {Maekawa}}]{yamashita2005}%
  \BibitemOpen
  \bibfield  {author} {\bibinfo {author} {\bibfnamefont {T.}~\bibnamefont
  {Yamashita}}, \bibinfo {author} {\bibfnamefont {K.}~\bibnamefont {Tanikawa}},
  \bibinfo {author} {\bibfnamefont {S.}~\bibnamefont {Takahashi}}, \ and\
  \bibinfo {author} {\bibfnamefont {S.}~\bibnamefont {Maekawa}},\ }\href@noop
  {} {\bibfield  {journal} {\bibinfo  {journal} {Phys. Rev. Lett.}\ }\textbf
  {\bibinfo {volume} {95}},\ \bibinfo {pages} {097001} (\bibinfo {year}
  {2005})}\BibitemShut {NoStop}%
\bibitem [{\citenamefont {Feofanov}\ \emph {et~al.}(2010)\citenamefont
  {Feofanov}, \citenamefont {Oboznov}, \citenamefont {Bol’ginov},
  \citenamefont {Lisenfeld}, \citenamefont {Poletto}, \citenamefont {Ryazanov},
  \citenamefont {Rossolenko}, \citenamefont {Khabipov}, \citenamefont
  {Balashov}, \citenamefont {Zorin}, \citenamefont {Dmitriev}, \citenamefont
  {Koshelets},\ and\ \citenamefont {Ustinov}}]{feofanov2010}%
  \BibitemOpen
  \bibfield  {author} {\bibinfo {author} {\bibfnamefont {A.~K.}\ \bibnamefont
  {Feofanov}}, \bibinfo {author} {\bibfnamefont {V.~A.}\ \bibnamefont
  {Oboznov}}, \bibinfo {author} {\bibfnamefont {V.~V.}\ \bibnamefont
  {Bol’ginov}}, \bibinfo {author} {\bibfnamefont {J.}~\bibnamefont
  {Lisenfeld}}, \bibinfo {author} {\bibfnamefont {S.}~\bibnamefont {Poletto}},
  \bibinfo {author} {\bibfnamefont {V.~V.}\ \bibnamefont {Ryazanov}}, \bibinfo
  {author} {\bibfnamefont {A.~N.}\ \bibnamefont {Rossolenko}}, \bibinfo
  {author} {\bibfnamefont {M.}~\bibnamefont {Khabipov}}, \bibinfo {author}
  {\bibfnamefont {D.}~\bibnamefont {Balashov}}, \bibinfo {author}
  {\bibfnamefont {A.~B.}\ \bibnamefont {Zorin}}, \bibinfo {author}
  {\bibfnamefont {P.~N.}\ \bibnamefont {Dmitriev}}, \bibinfo {author}
  {\bibfnamefont {V.~P.}\ \bibnamefont {Koshelets}}, \ and\ \bibinfo {author}
  {\bibfnamefont {A.~V.}\ \bibnamefont {Ustinov}},\ }\href@noop {} {\bibfield
  {journal} {\bibinfo  {journal} {Nat. Phys.}\ }\textbf {\bibinfo {volume}
  {6}},\ \bibinfo {pages} {593} (\bibinfo {year} {2010})}\BibitemShut {NoStop}%
\bibitem [{\citenamefont {Hinton}\ \emph {et~al.}(2016)\citenamefont {Hinton},
  \citenamefont {Steers}, \citenamefont {Peters}, \citenamefont {Yang},\ and\
  \citenamefont {Lemberger}}]{hinton2016}%
  \BibitemOpen
  \bibfield  {author} {\bibinfo {author} {\bibfnamefont {M.~J.}\ \bibnamefont
  {Hinton}}, \bibinfo {author} {\bibfnamefont {S.}~\bibnamefont {Steers}},
  \bibinfo {author} {\bibfnamefont {B.}~\bibnamefont {Peters}}, \bibinfo
  {author} {\bibfnamefont {F.~Y.}\ \bibnamefont {Yang}}, \ and\ \bibinfo
  {author} {\bibfnamefont {T.~R.}\ \bibnamefont {Lemberger}},\ }\href@noop {}
  {\bibfield  {journal} {\bibinfo  {journal} {Phys. Rev. B}\ }\textbf {\bibinfo
  {volume} {94}},\ \bibinfo {pages} {014518} (\bibinfo {year}
  {2016})}\BibitemShut {NoStop}%
\bibitem [{\citenamefont {Torokhtii}\ \emph {et~al.}(2013)\citenamefont
  {Torokhtii}, \citenamefont {Pompeo}, \citenamefont {Meneghini}, \citenamefont
  {Attanasio}, \citenamefont {Cirillo}, \citenamefont {amd S.~Sarti},\ and\
  \citenamefont {Silva}}]{Pompeo}%
  \BibitemOpen
  \bibfield  {author} {\bibinfo {author} {\bibfnamefont {K.}~\bibnamefont
  {Torokhtii}}, \bibinfo {author} {\bibfnamefont {N.}~\bibnamefont {Pompeo}},
  \bibinfo {author} {\bibfnamefont {C.}~\bibnamefont {Meneghini}}, \bibinfo
  {author} {\bibfnamefont {C.}~\bibnamefont {Attanasio}}, \bibinfo {author}
  {\bibfnamefont {C.}~\bibnamefont {Cirillo}}, \bibinfo {author} {\bibfnamefont
  {E.~A.~I.}\ \bibnamefont {amd S.~Sarti}}, \ and\ \bibinfo {author}
  {\bibfnamefont {E.}~\bibnamefont {Silva}},\ }\href@noop {} {\bibfield
  {journal} {\bibinfo  {journal} {J Supercond. Nov. Magn.}\ }\textbf {\bibinfo
  {volume} {26}},\ \bibinfo {pages} {571} (\bibinfo {year} {2013})}\BibitemShut
  {NoStop}%
\bibitem [{\citenamefont {Schmid}\ and\ \citenamefont
  {Sch\"on}(1975)}]{Schmid1975}%
  \BibitemOpen
  \bibfield  {author} {\bibinfo {author} {\bibfnamefont {A.}~\bibnamefont
  {Schmid}}\ and\ \bibinfo {author} {\bibfnamefont {G.}~\bibnamefont
  {Sch\"on}},\ }\href@noop {} {\bibfield  {journal} {\bibinfo  {journal} {J.
  Low Temp. Phys.}\ }\textbf {\bibinfo {volume} {20}},\ \bibinfo {pages} {207}
  (\bibinfo {year} {1975})}\BibitemShut {NoStop}%
\bibitem [{\citenamefont {Bergeret}\ \emph {et~al.}(2018)\citenamefont
  {Bergeret}, \citenamefont {Silaev}, \citenamefont {Virtanen},\ and\
  \citenamefont {Heikkil\"a}}]{bergeret2018}%
  \BibitemOpen
  \bibfield  {author} {\bibinfo {author} {\bibfnamefont {F.~S.}\ \bibnamefont
  {Bergeret}}, \bibinfo {author} {\bibfnamefont {M.}~\bibnamefont {Silaev}},
  \bibinfo {author} {\bibfnamefont {P.}~\bibnamefont {Virtanen}}, \ and\
  \bibinfo {author} {\bibfnamefont {T.~T.}\ \bibnamefont {Heikkil\"a}},\
  }\href@noop {} {\bibfield  {journal} {\bibinfo  {journal} {Rev. Mod. Phys.}\
  }\textbf {\bibinfo {volume} {90}},\ \bibinfo {pages} {041001} (\bibinfo
  {year} {2018})}\BibitemShut {NoStop}%
\bibitem [{\citenamefont {Virtanen}\ \emph {et~al.}(2019)\citenamefont
  {Virtanen}, \citenamefont {Vargunin},\ and\ \citenamefont
  {Silaev}}]{1909.00992}%
  \BibitemOpen
  \bibfield  {author} {\bibinfo {author} {\bibfnamefont {P.}~\bibnamefont
  {Virtanen}}, \bibinfo {author} {\bibfnamefont {A.}~\bibnamefont {Vargunin}},
  \ and\ \bibinfo {author} {\bibfnamefont {M.}~\bibnamefont {Silaev}},\
  }\href@noop {} {\enquote {\bibinfo {title} {Quasiclassical expressions for
  the free energy of superconducting systems},}\ } (\bibinfo {year} {2019}),\
  \Eprint {http://arxiv.org/abs/arXiv:1909.00992} {arXiv:1909.00992}
  \BibitemShut {NoStop}%
\bibitem [{SM()}]{SM}%
  \BibitemOpen
  \href@noop {} {}\bibinfo {note} {See Supplementary Material for
  details}\BibitemShut {NoStop}%
\bibitem [{\citenamefont {Samokhvalov}\ and\ \citenamefont
  {Buzdin}(2015)}]{samokhvalov2015}%
  \BibitemOpen
  \bibfield  {author} {\bibinfo {author} {\bibfnamefont {A.~V.}\ \bibnamefont
  {Samokhvalov}}\ and\ \bibinfo {author} {\bibfnamefont {A.~I.}\ \bibnamefont
  {Buzdin}},\ }\href@noop {} {\bibfield  {journal} {\bibinfo  {journal} {Phys.
  Rev. B}\ }\textbf {\bibinfo {volume} {92}},\ \bibinfo {pages} {054511}
  (\bibinfo {year} {2015})}\BibitemShut {NoStop}%
\bibitem [{\citenamefont {Fominov}\ \emph {et~al.}(2002)\citenamefont
  {Fominov}, \citenamefont {Chtchelkatchev},\ and\ \citenamefont
  {Golubov}}]{Fominov2002}%
  \BibitemOpen
  \bibfield  {author} {\bibinfo {author} {\bibfnamefont {Y.~V.}\ \bibnamefont
  {Fominov}}, \bibinfo {author} {\bibfnamefont {N.~M.}\ \bibnamefont
  {Chtchelkatchev}}, \ and\ \bibinfo {author} {\bibfnamefont {A.~A.}\
  \bibnamefont {Golubov}},\ }\href@noop {} {\bibfield  {journal} {\bibinfo
  {journal} {Phys. Rev. B}\ }\textbf {\bibinfo {volume} {66}},\ \bibinfo
  {pages} {014507} (\bibinfo {year} {2002})}\BibitemShut {NoStop}%
\bibitem [{\citenamefont {Krunavakarn}\ and\ \citenamefont
  {Yoksan}(2006)}]{Krunavakarn2006}%
  \BibitemOpen
  \bibfield  {author} {\bibinfo {author} {\bibfnamefont {B.}~\bibnamefont
  {Krunavakarn}}\ and\ \bibinfo {author} {\bibfnamefont {S.}~\bibnamefont
  {Yoksan}},\ }\href@noop {} {\bibfield  {journal} {\bibinfo  {journal}
  {Physica C}\ }\textbf {\bibinfo {volume} {440}},\ \bibinfo {pages} {25}
  (\bibinfo {year} {2006})}\BibitemShut {NoStop}%
\bibitem [{\citenamefont {Jiang}\ \emph {et~al.}(1995)\citenamefont {Jiang},
  \citenamefont {Davidovic}, \citenamefont {Reich},\ and\ \citenamefont
  {Chien}}]{jiang1995}%
  \BibitemOpen
  \bibfield  {author} {\bibinfo {author} {\bibfnamefont {J.}~\bibnamefont
  {Jiang}}, \bibinfo {author} {\bibfnamefont {D.}~\bibnamefont {Davidovic}},
  \bibinfo {author} {\bibfnamefont {D.~H.}\ \bibnamefont {Reich}}, \ and\
  \bibinfo {author} {\bibfnamefont {C.~L.}\ \bibnamefont {Chien}},\ }\href@noop
  {} {\bibfield  {journal} {\bibinfo  {journal} {Phys. Rev. Lett.}\ }\textbf
  {\bibinfo {volume} {74}},\ \bibinfo {pages} {314} (\bibinfo {year}
  {1995})}\BibitemShut {NoStop}%
\bibitem [{\citenamefont {Samokhvalov}(2017)}]{samokhvalov2017}%
  \BibitemOpen
  \bibfield  {author} {\bibinfo {author} {\bibfnamefont {A.~V.}\ \bibnamefont
  {Samokhvalov}},\ }\href@noop {} {\bibfield  {journal} {\bibinfo  {journal}
  {Phys. Solid State}\ }\textbf {\bibinfo {volume} {59}},\ \bibinfo {pages}
  {2143} (\bibinfo {year} {2017})}\BibitemShut {NoStop}%
\bibitem [{\citenamefont {Golubov}\ \emph {et~al.}(2016)\citenamefont
  {Golubov}, \citenamefont {Kupriyanov},\ and\ \citenamefont
  {Khapaev}}]{golubov2016abrikosov}%
  \BibitemOpen
  \bibfield  {author} {\bibinfo {author} {\bibfnamefont {A.~A.}\ \bibnamefont
  {Golubov}}, \bibinfo {author} {\bibfnamefont {M.~Y.}\ \bibnamefont
  {Kupriyanov}}, \ and\ \bibinfo {author} {\bibfnamefont {M.~M.}\ \bibnamefont
  {Khapaev}},\ }\href@noop {} {\bibfield  {journal} {\bibinfo  {journal} {JETP
  letters}\ }\textbf {\bibinfo {volume} {104}},\ \bibinfo {pages} {847}
  (\bibinfo {year} {2016})}\BibitemShut {NoStop}%
\bibitem [{\citenamefont {Stolyarov}\ \emph {et~al.}(2018)\citenamefont
  {Stolyarov}, \citenamefont {Cren}, \citenamefont {Brun}, \citenamefont
  {Golovchanskiy}, \citenamefont {Skryabina}, \citenamefont {Kasatonov},
  \citenamefont {Khapaev}, \citenamefont {Kupriyanov}, \citenamefont
  {Golubov},\ and\ \citenamefont {Roditchev}}]{stolyarov2018expansion}%
  \BibitemOpen
  \bibfield  {author} {\bibinfo {author} {\bibfnamefont {V.~S.}\ \bibnamefont
  {Stolyarov}}, \bibinfo {author} {\bibfnamefont {T.}~\bibnamefont {Cren}},
  \bibinfo {author} {\bibfnamefont {C.}~\bibnamefont {Brun}}, \bibinfo {author}
  {\bibfnamefont {I.~A.}\ \bibnamefont {Golovchanskiy}}, \bibinfo {author}
  {\bibfnamefont {O.~V.}\ \bibnamefont {Skryabina}}, \bibinfo {author}
  {\bibfnamefont {D.~I.}\ \bibnamefont {Kasatonov}}, \bibinfo {author}
  {\bibfnamefont {M.~M.}\ \bibnamefont {Khapaev}}, \bibinfo {author}
  {\bibfnamefont {M.~Y.}\ \bibnamefont {Kupriyanov}}, \bibinfo {author}
  {\bibfnamefont {A.~A.}\ \bibnamefont {Golubov}}, \ and\ \bibinfo {author}
  {\bibfnamefont {D.}~\bibnamefont {Roditchev}},\ }\href@noop {} {\bibfield
  {journal} {\bibinfo  {journal} {Nature communications}\ }\textbf {\bibinfo
  {volume} {9}},\ \bibinfo {pages} {2277} (\bibinfo {year} {2018})}\BibitemShut
  {NoStop}%
\bibitem [{\citenamefont {Ihle}(1971)}]{ihle1971b}%
  \BibitemOpen
  \bibfield  {author} {\bibinfo {author} {\bibfnamefont {D.}~\bibnamefont
  {Ihle}},\ }\href@noop {} {\bibfield  {journal} {\bibinfo  {journal} {Phys.
  Status Solidi B}\ }\textbf {\bibinfo {volume} {47}},\ \bibinfo {pages} {429}
  (\bibinfo {year} {1971})}\BibitemShut {NoStop}%
\bibitem [{\citenamefont {Watts-Tobin}\ \emph {et~al.}(1974)\citenamefont
  {Watts-Tobin}, \citenamefont {Kramer},\ and\ \citenamefont
  {Pesch}}]{WattsTobin1974}%
  \BibitemOpen
  \bibfield  {author} {\bibinfo {author} {\bibfnamefont {R.~J.}\ \bibnamefont
  {Watts-Tobin}}, \bibinfo {author} {\bibfnamefont {L.}~\bibnamefont {Kramer}},
  \ and\ \bibinfo {author} {\bibfnamefont {W.}~\bibnamefont {Pesch}},\
  }\href@noop {} {\bibfield  {journal} {\bibinfo  {journal} {J. Low Temp.
  Phys.}\ }\textbf {\bibinfo {volume} {17}},\ \bibinfo {pages} {71} (\bibinfo
  {year} {1974})}\BibitemShut {NoStop}%
\bibitem [{\citenamefont {Rammer}\ \emph {et~al.}(1987)\citenamefont {Rammer},
  \citenamefont {Pesch},\ and\ \citenamefont {Kramer}}]{Rammer1987}%
  \BibitemOpen
  \bibfield  {author} {\bibinfo {author} {\bibfnamefont {J.}~\bibnamefont
  {Rammer}}, \bibinfo {author} {\bibfnamefont {W.}~\bibnamefont {Pesch}}, \
  and\ \bibinfo {author} {\bibfnamefont {L.}~\bibnamefont {Kramer}},\
  }\href@noop {} {\bibfield  {journal} {\bibinfo  {journal} {Z. Phys. B}\
  }\textbf {\bibinfo {volume} {68}},\ \bibinfo {pages} {49} (\bibinfo {year}
  {1987})}\BibitemShut {NoStop}%
\bibitem [{\citenamefont {Rammer}(1988)}]{Rammer1988}%
  \BibitemOpen
  \bibfield  {author} {\bibinfo {author} {\bibfnamefont {J.}~\bibnamefont
  {Rammer}},\ }\href@noop {} {\bibfield  {journal} {\bibinfo  {journal} {J. Low
  Temp. Phys.}\ }\textbf {\bibinfo {volume} {71}},\ \bibinfo {pages} {323}
  (\bibinfo {year} {1988})}\BibitemShut {NoStop}%
\bibitem [{\citenamefont {Larkin}\ and\ \citenamefont
  {Ovchinnikov}(1986)}]{LO}%
  \BibitemOpen
  \bibfield  {author} {\bibinfo {author} {\bibfnamefont {A.~I.}\ \bibnamefont
  {Larkin}}\ and\ \bibinfo {author} {\bibfnamefont {Y.~N.}\ \bibnamefont
  {Ovchinnikov}},\ }\href@noop {} {\emph {\bibinfo {title} {Modern Problems in
  Condensed Matter Sciences: Nonequilibrium Superconductivity}}},\ edited by\
  \bibinfo {editor} {\bibfnamefont {D.~N.}\ \bibnamefont {Langenberg}}\ and\
  \bibinfo {editor} {\bibfnamefont {A.~I.}\ \bibnamefont {Larkin}}\ (\bibinfo
  {publisher} {Elsevier},\ \bibinfo {year} {1986})\ p.\ \bibinfo {pages}
  {493}\BibitemShut {NoStop}%
\bibitem [{\citenamefont {Kopnin}(2001)}]{kopninbook}%
  \BibitemOpen
  \bibfield  {author} {\bibinfo {author} {\bibfnamefont {N.~B.}\ \bibnamefont
  {Kopnin}},\ }\href@noop {} {\emph {\bibinfo {title} {Theory of Nonequilibrium
  Superconductivity}}}\ (\bibinfo  {publisher} {Oxford University Press},\
  \bibinfo {year} {2001})\BibitemShut {NoStop}%
\bibitem [{\citenamefont {Mel'Nikov}(1996)}]{mel1996inertial}%
  \BibitemOpen
  \bibfield  {author} {\bibinfo {author} {\bibfnamefont {A.~S.}\ \bibnamefont
  {Mel'Nikov}},\ }\href@noop {} {\bibfield  {journal} {\bibinfo  {journal}
  {Phys. Rev. Lett.}\ }\textbf {\bibinfo {volume} {77}},\ \bibinfo {pages}
  {2786} (\bibinfo {year} {1996})}\BibitemShut {NoStop}%
\bibitem [{\citenamefont {Radovi\'c}\ \emph {et~al.}(1991)\citenamefont
  {Radovi\'c}, \citenamefont {Ledvij},\ and\ \citenamefont
  {Dobrosavljevi\'c-Gruji\'c}}]{radovic1991}%
  \BibitemOpen
  \bibfield  {author} {\bibinfo {author} {\bibfnamefont {Z.}~\bibnamefont
  {Radovi\'c}}, \bibinfo {author} {\bibfnamefont {M.}~\bibnamefont {Ledvij}}, \
  and\ \bibinfo {author} {\bibfnamefont {L.}~\bibnamefont
  {Dobrosavljevi\'c-Gruji\'c}},\ }\href@noop {} {\bibfield  {journal} {\bibinfo
   {journal} {Solid State Commun.}\ }\textbf {\bibinfo {volume} {80}},\
  \bibinfo {pages} {43} (\bibinfo {year} {1991})}\BibitemShut {NoStop}%
\bibitem [{\citenamefont {Vargunin}\ and\ \citenamefont {Silaev}(2017)}]{vs}%
  \BibitemOpen
  \bibfield  {author} {\bibinfo {author} {\bibfnamefont {A.}~\bibnamefont
  {Vargunin}}\ and\ \bibinfo {author} {\bibfnamefont {M.~A.}\ \bibnamefont
  {Silaev}},\ }\href@noop {} {\bibfield  {journal} {\bibinfo  {journal} {Phys.
  Rev. B}\ }\textbf {\bibinfo {volume} {96}},\ \bibinfo {pages} {214507}
  (\bibinfo {year} {2017})}\BibitemShut {NoStop}%
\end{thebibliography}%

\end{document}